\documentclass[
reprint,
superscriptaddress,
 amsmath,amssymb,
pra,
]{revtex4-2}

\usepackage{graphicx}
\usepackage{dcolumn}
\usepackage{bm}

\begin{document}

\preprint{APS/123-QED}

\title{Oxygen Vacancies in Niobium Pentoxide as a Source of Two-Level System \\ Losses in Superconducting Niobium}

\author{D. Bafia}
 \altaffiliation{dbafia@fnal.gov}
 \affiliation{
 Fermi National Accelerator Laboratory, Batavia, Illinois 60510, USA}
\author{A. Murthy}
\affiliation{
 Fermi National Accelerator Laboratory, Batavia, Illinois 60510, USA}
\author{A. Grassellino}
\affiliation{
 Fermi National Accelerator Laboratory, Batavia, Illinois 60510, USA}
\author{A. Romanenko}
\affiliation{
 Fermi National Accelerator Laboratory, Batavia, Illinois 60510, USA}

\date{\today}

\begin{abstract}

We identify a major source of quantum decoherence in three-dimensional superconducting radio-frequency (SRF) resonators and two-dimensional transmon qubits composed of oxidized niobium: oxygen vacancies in the niobium pentoxide which drive two-level system (TLS) losses. By probing the effect of sequential \textit{in situ} vacuum baking treatments on the RF performance of bulk Nb SRF resonators and on the oxide structure of a representative Nb sample using time-of-flight secondary ion mass spectrometry (ToF-SIMS), we find a non-monotonic evolution of cavity quality factor $Q_0$ which correlates with the interplay of Nb\textsubscript{2}O\textsubscript{5} vacancy generation and oxide thickness reduction. We localize this effect to the oxide itself and present the insignificant role of diffused interstitial oxygen in the underlying Nb by regrowing a new oxide \textit{via} wet oxidation which reveals a mitigation of aggravated TLS losses. We hypothesize that such vacancies in the pentoxide serve as magnetic impurities and are a source of TLS-driven RF loss.

\end{abstract}

\maketitle

\section{Introduction\label{sec:Intro}}
 
The realization of a computationally useful superconducting quantum computer requires the maximization of quantum coherence lifetimes to unlock quantum memory, high-fidelity gates, and better entanglement among qubits \cite{Clarke_Nature_2008}. In superconducting systems, quantum coherence is limited by dissipative decay channels present in both the linear resonator and the transmon qubit that contains the nonlinear Josephson junction necessary to create uniquely addressable energy levels. Current decoherence candidates include quasi-particles \cite{Graaf_arXiv_2020, DeVisser_PRL_2014}, radiation-induced losses \cite{Veps_arXiv_2020, Martinis_Nature_2021}, and dielectric losses such as two-level systems (TLS) \cite{Pappas_IEEE_2011}, to name a few. Dielectric loss arises from each material and interface used in the construction of superconducting qubits and dictates the total decay rate of the system according to

\begin{equation}
    \Gamma _1 = \frac{1}{T_1} = \frac{\omega}{Q _0} = \omega\sum_{i} \frac{F_i}{Q_i} + \Gamma _0,
    \label{eq:T1}
\end{equation}

\noindent where $T_1$ is the photon lifetime, $\omega$ is angular frequency, $Q_0$ is the quality factor, $F_i$ and $Q_i$ are the filling and quality factors of the $i^{th}$ loss channel, and $\Gamma _0$ captures losses driven by non-dielectric mechanisms \cite{Wang_APL_2015,Woods_PRA_2019}. The quality factor is inversely related to the dielectric loss tangent $Q_i = 1/\text{tan} \delta _i$. As a result, decay rate minimization requires the mitigation of every loss channel down to tolerable levels. While straightforward for a single interface, this task is difficult in complex multi-layer systems like those used in modern qubit fabrication designs. Many such designs utilize thin films of aluminum and niobium atop dielectric or semiconducting substrates, leading to multiple potentially deleterious materials and interfaces.

Recent work on three-dimensional (3-D) niobium superconducting radio-frequency (SRF) resonators in the low GHz frequency range has highlighted the amorphous native niobium oxide as a dramatic source of TLS loss, implying an inherent limitation in qubits fabricated with oxidized niobium \cite{Rom17,Rom20}. By removing the native niobium oxide \textit{via} \textit{in situ} vacuum baking at 340-450$^{\circ}$C for several hours, a ten-fold improvement in quality factor $Q_0$ was observed, demonstrating intra-resonator photon lifetimes of $T_1\approx$~2~s, a 200-fold increase over the previous state-of-the-art \cite{Reagor_APL_2013}. By fitting temperature dependent $Q_0$ data of bulk niobium resonators with the standard TLS model \cite{Rom17,Rom20}, Romanenko \textit{et al.} found the dielectric loss tangent of the native niobium oxide at 5~GHz to be $\delta_0$~=~0.08. This value is in agreement with values obtained from niobium-on-silicon lumped element resonators post sequential oxidation \cite{Verjauw_PRA_2021} and niobium-on-silicon coplanar-waveguides post sequential etching \cite{Altoe_PRX_2022}. This consistency between dramatically different architectures demonstrates that observations of the Nb-air interface made in simplified, single interface systems such as 3-D Nb SRF resonators are transferable to more complex systems such as those used in transmon qubits, allowing for a one-to-one correspondence between thin film and bulk material measurements.

While there exists an empirical mitigation strategy to minimize their deleterious effect in superconducting quantum systems, the precise origin and nature of the TLS losses within the niobium oxide is not yet fully understood. A complete picture of the underlying mechanisms would aid in the further development of mitigation strategies that may instead be stable to air. There exist a few possible origins for these TLS losses. One possibility stems from the conducting or magnetic nature of the NbO\textsubscript{x} compounds \cite{Rom20}. Another lies in the theory that interstitial oxygen in the niobium lattice could provide tunneling sites for hydrogen \cite{Wipf_PRL_1984}. A third possibility involves oxygen vacancies within the amorphous pentoxide acting as magnetic impurities \cite{Cava_PRB_1991,Proslier_2008} and driving impedance, as supported by X-ray absorption spectroscopy studies \cite{Harrelson_APL_2021} and density functional theory calculations \cite{Harrelson_APL_2021,Sheridan_arXiv_2021}. Here, we present results that support the role of the latter in driving TLS losses in oxidized niobium.

One avenue for identifying the origins of loss in the amorphous native Nb oxide is by studying its evolution with \textit{in situ} baking, allowing for the investigation of the role of various oxides on $Q_0$. There exists extensive literature on this topic due to its relevance in high $Q_0$ and high electric field applications of Nb SRF resonators in particle accelerators. It is well known that native niobium oxides exist in a graded structure composed of Nb\textsubscript{2}O\textsubscript{5} (insulator \cite{Halbritter_APL_1987}), NbO\textsubscript{2} (semiconductor \cite{Zhao_2004}), NbO (metallic \cite{Dacca_ASS_1998}), and the metal substrate \cite{Halbritter_APL_1987, King_1990,Murthy_ACS_2022}. \textit{In situ} vacuum baking dissolves the dominant pentoxide and grows the suboxides, reducing according to Nb\textsubscript{2}O\textsubscript{5}~$\rightarrow$~NbO\textsubscript{2}~$\rightarrow$~NbO \cite{King_1990, Ma_JAP_2004, Del_APL_2008, MA_ASS_2003}. Ma \textit{et al.} showed that \textit{in situ} vacuum baking at 150$^{\circ}$C for 40~hours reduced the pentoxide to NbO\textsubscript{2}; continued baking at 280$^{\circ}$C for 8~hours showed further reduction to metallic Nb\textsubscript{x}O (x$\sim$2) \cite{Ma_JAP_2004}. Further studies by Ma \textit{et al.} revealed the reduction of the pentoxide to NbO\textsubscript{2} and NbO post vacuum baking at 250$^{\circ}$C for several hours \cite{MA_ASS_2003}. These results are corroborated by the synchrotron measurements of Delheusy \textit{et al.} \cite{Del_APL_2008} and ARXPS measurements by Dacca \textit{et al.} \cite{Dacca_ASS_1998}. Moreover, at the expense of dissolving the oxide, \textit{in situ} vacuum baking drives inward oxygen diffusion from the oxide over the oxide/metal interface rather than into the vacuum \cite{Ma_JAP_2004, Schulze_1977, Fromm_1977, Del_APL_2008, Lechner_APL_2021}. Subsequent reoxidation studies \textit{via} exposure to ambient air shows that a partially dissolved oxide grows at the oxide/metal interface and is morphologically equivalent to the native grown oxide without prior heat treatment \cite{MA_ASS_2003}. 

In this paper, by investigating the role of vacuum baking on niobium SRF resonators which contain a $\sim$5~nm full wet-grown native amorphous oxide, we find that vacancies in the niobium pentoxide host TLS which limit the performance of 3-D Nb resonators and 2-D transmon qubits which utilize oxidized Nb. Modest \textit{in situ} vacuum baking treatments (150$^{\circ}$C-200$^{\circ}$C) for durations as short as 25~minutes aggravate TLS-like losses that are eventually suppressed with increasing bake time. We recreate the \textit{in situ} vacuum baking treatments using time-of-flight secondary ion mass spectrometry (ToF-SIMS) on a Nb SRF resonator cutout and confirm that the oxide reduces considerably, suggesting an increase in the number of oxygen vacancies in the Nb\textsubscript{2}O\textsubscript{5} layer. Continued \textit{in situ} baking shows that the total number of oxygen vacancies in the pentoxide decreases as the layer is gradually reduced into suboxides. As such, the non-monotonic evolution of $Q_0$ with successive baking is due to the interplay of oxygen vacancy generation in the Nb\textsubscript{2}O\textsubscript{5} layer, which host TLS, and subsequent pentoxide dissolution. Finally, we consider one potential model in which these oxygen vacancies drive TLS, namely, that they host magnetic impurities.

\section{Experimental Method\label{sec:Experimental}}

For our studies, we used three 1.3 GHz TESLA-shaped \cite{Aune00} niobium single-cell SRF resonators that have undergone a bulk 120~$\mu$m removal from the inner RF surface \textit{via} electropolishing, followed by 800$^{\circ}$C degassing, and an additional 40~$\mu$m removal from the inner surface \textit{via} electropolishing \cite{Padamsee98}. The resonators were then exposed to air and high pressure rinsed with ultra-pure water, forming a $\sim$5~nm amorphous native oxide. Afterwards, the resonators were assembled for testing and evacuated to a vacuum level of approximately 1E-5~Torr. Typically, resonators tested at this point are called electropolished (EP) resonators. The resonators then underwent sequential testing and treatment to various low temperature vacuum bakes  \cite{Padamsee09} for increasing durations while actively pumping to maintain the vacuum level within the resonators. Such treatments have been shown to dissolve the native niobium oxide and introduce oxygen interstitial into the underlying niobium \cite{Ma_JAP_2004, Schulze_1977, Fromm_1977, Del_APL_2008, Bafia_SRF_2021, Lechner_APL_2021}.

One resonator, TE1AES019, was first subjected to an \textit{in situ} low temperature vacuum bake at 90$^{\circ}$C for 384~hours followed by sequential \textit{in situ} baking steps at 200$^{\circ}$C for increasing durations. For the final treatment, the resonator was baked at 340$^{\circ}$C for 5 hours to ensure complete dissolution of the niobium pentoxide \cite{Rom20}. The resonator was RF tested after each step and maintained vacuum throughout the entire course of the study. 

The second resonator, TE1RI003, was used to isolate losses to the dynamics of the oxide itself. After its baseline test post electropolishing, the cavity underwent \textit{in situ} low temperature baking at 200$^{\circ}$C for 1 hour followed by further testing. The cavity was then vented to clean room air and left to oxidize for a full week. This drove the re-formation of the native niobium oxide but this time with a larger concentration of subsurface oxygen (O) interstitial present compared to the baseline measurement post EP. Once complete, TE1RI003 was once again re-evacuated and tested.

The third resonator, TE1AES021, was used to probe baking treatments that more closely resembled those typically utilized in transmon qubit fabrication and to further identify the source of additional RF loss. It underwent sequential \textit{in situ} vacuum baking treatments at 150$^{\circ}$C with resonator RF measurements performed after each step. After the RF test post baking at 150$^{\circ}$C for a net duration of 15 hours and 25 minutes, resonator TE1AES021 was subjected to an \textit{ex situ} vacuum bake at 200$^{\circ}$C for 19.5 hours. This involved first venting the resonator to atmosphere followed by vacuum baking. Afterwards, the resonator was once again high pressure rinsed, reforming the amorphous $\sim$5~nm native oxide, again increasing the concentration of subsurface interstitial O compared to the baseline EP treatment. For the final treatment, the resonator underwent an HF acid rinse, which dissolved the native oxide, and upon subsequent high pressure rinsing, reformed a new native oxide at the expense of consuming approximately 2~nm of niobium \cite{Checchin_APL_2020}. 

To test our resonators, we utilized methods similar to those outlined in \cite{Rom17,Melnychuk_RSI_2014}. Resonators were installed in large vertical helium dewars and cooled through the superconducting transition temperature at $\sim$9.25~K while employing methods to minimize the possibility of trapping magnetic flux \cite{Romanenko_JAP_2014,Posen_JAP_2016}. Resonators were then driven at their resonant frequency while using a phase-locked loop to track and maintain resonance. We used a signal analyzer set to a resolution bandwidth of $\sim$220~Hz to perform single shot, zero span decay measurements of the transmitted power $P_t (t)$ from the resonator after shutting off the incident RF power. By fitting the resulting $P_t (t)$ data with the procedure outlined in \cite{Rom17}, we obtained a measure of $Q_0$ down to fields of $E_{acc}\sim$~1~kV/m with experimental error never exceeding 10~$\%$. The fields were then converted into a measure of the photon number \textit{via} $n=U/\hbar \omega$, where $U$ is the stored energy in the resonator, which is related to $P_t (t)$. We performed the tests at temperatures of 1.4-1.6~K, where the population of quasi-particles was virtually non-existent, and their contribution to the surface resistance was negligible. As a result, the quality factor in these tests was dominated by losses stemming from material properties. We note that while most TLS remain thermally saturated at these temperatures, there remains a small fraction of unsaturated TLS which drive the characteristic response of $Q_0$ with intra-resonator photon number \cite{Rom17,Rom20}.

\section{Results}
\subsection{Resonator RF measurements}

\begin{figure}
    \centering
    \includegraphics[width=8.2cm]{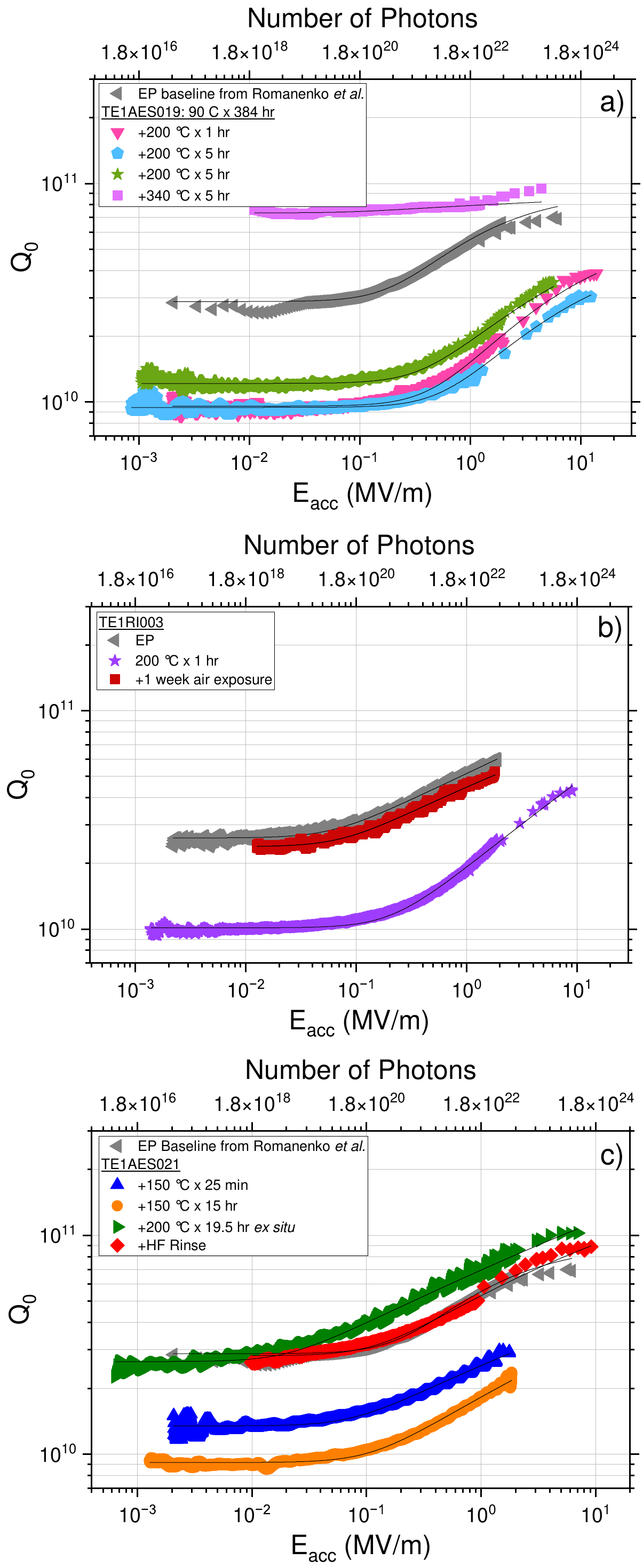}
\caption{Measurements of quality factor vs electric field at the center of the cavity at temperatures of 1.4~K-1.6~K for the 1.3 GHz 3-D Nb resonators a) TE1AES019, b) TE1RI003, and c) TE1AES021 after sequential baking and chemical treatments. Upper axes plot the converted intra-resonator photon number. For reference, the gray left pointed triangles in a) and c) depict data obtained on an EP resonator which contained a full 5~nm native wet grown oxide from Romanenko and Schuster \cite{Rom17}. Solid black lines show fits with Eq.~\ref{eq:PowDepTLS}.}
    \label{fig:QvsE}
\end{figure}

The upper panel in Fig.~\ref{fig:QvsE} presents data acquired on resonator TE1AES019 after an initial 90$^{\circ}$C~$\times$~384~hour \textit{in situ} vacuum bake followed by sequential \textit{in situ} vacuum bakes at various temperatures and durations. Shown also for reference is data from Romanenko and Schuster taken on an EP resonator which contains a full 5~nm native wet-grown oxide and serves as our baseline test \cite{Rom17}. For all tests, we observe the expected saturation of the quality factor at electric fields $<$~0.01~MV/m (i.e. photon numbers $<$~2E18) that is characteristic of two-level systems \cite{Rom17, Rom20, Pappas_IEEE_2011}. We find that the 90$^{\circ}$C~$\times$~384~hours~$+$~200$^{\circ}$C~$\times$~1~hour \textit{in situ} vacuum baking treatment produces a resonator with a low field quality factor of 9E9, roughly three times lower than the baseline test. An additional 200$^{\circ}$C~$\times$~5~hour \textit{in situ} vacuum bake produces identical results at fields below 0.1~MV/m. After the third round of treatment of 200$^{\circ}$C~$\times$~5~hour vacuum bake, a slight improvement in $Q_0$ up to 1.2E10 occurs. For the final treatment, we subjected the resonator to a 340$^{\circ}$C~$\times$~5~hour vacuum baking treatment which is known to fully dissolve the native Nb\textsubscript{2}O\textsubscript{5} layer \cite{Rom20, Posen_PRA_2020}. The resulting measurements display a $Q_0$ at low fields of greater than 7E10, the highest achieved in this study, and in line with previous work \cite{Rom20, Posen_PRA_2020}. Overall, these tests show a non-monotonic dependence of resonator $Q_0$ with bake duration.

Fig.~\ref{fig:QvsE}b) instead plots the results of cavity TE1RI003. The baseline test post EP yields an expected low field $Q_0$ of 2.5E10. Post 200$^{\circ}$C~$\times$~1~hour \textit{in situ} baking, we observe a reduction of this value by a factor of approximately three, in agreement with that observed in Fig.~\ref{fig:QvsE}a). Subsequent re-exposure to clean room air for one week and further testing showed a mitigation of aggravated TLS losses and a return to EP baseline performance. These results highlight the role of oxide dynamics in driving cavity performance.

The lower panel in Fig.~\ref{fig:QvsE} depicts data taken on resonator TE1AES021 post various baking and chemical treatments. Surprisingly, we find that a modest \textit{in situ} vacuum bake at 150$^{\circ}$C for 25~minutes reduces the resonator $Q_0$ by a factor of two when compared to the baseline EP test from Romanenko and Schuster. Subjecting the resonator to an additional round of \textit{in situ} vacuum baking at 150$^{\circ}$C for 15~hours produces a low field $Q_0$ of 9E9, identical to the results obtained on TE1AES019 and TE1RI003 post  200$^{\circ}$C~$\times$~1~hour \textit{in situ} baking. The results post \textit{ex situ} bake at 200$^{\circ}$C for 19.5~hours, for which we re-iterate received subsequent high pressure rinsing with ultra-pure water, yield a $Q_0$ that is similar in value to the baseline EP at fields $<$~4E-3~MV/m. Subsequent HF acid rinsing and high pressure water rinsing reveals a $Q_0$ vs $E_{acc}$ curve that is nearly identical to the baseline EP data.

\begin{figure}
    \centering
    \includegraphics[width=8.6cm]{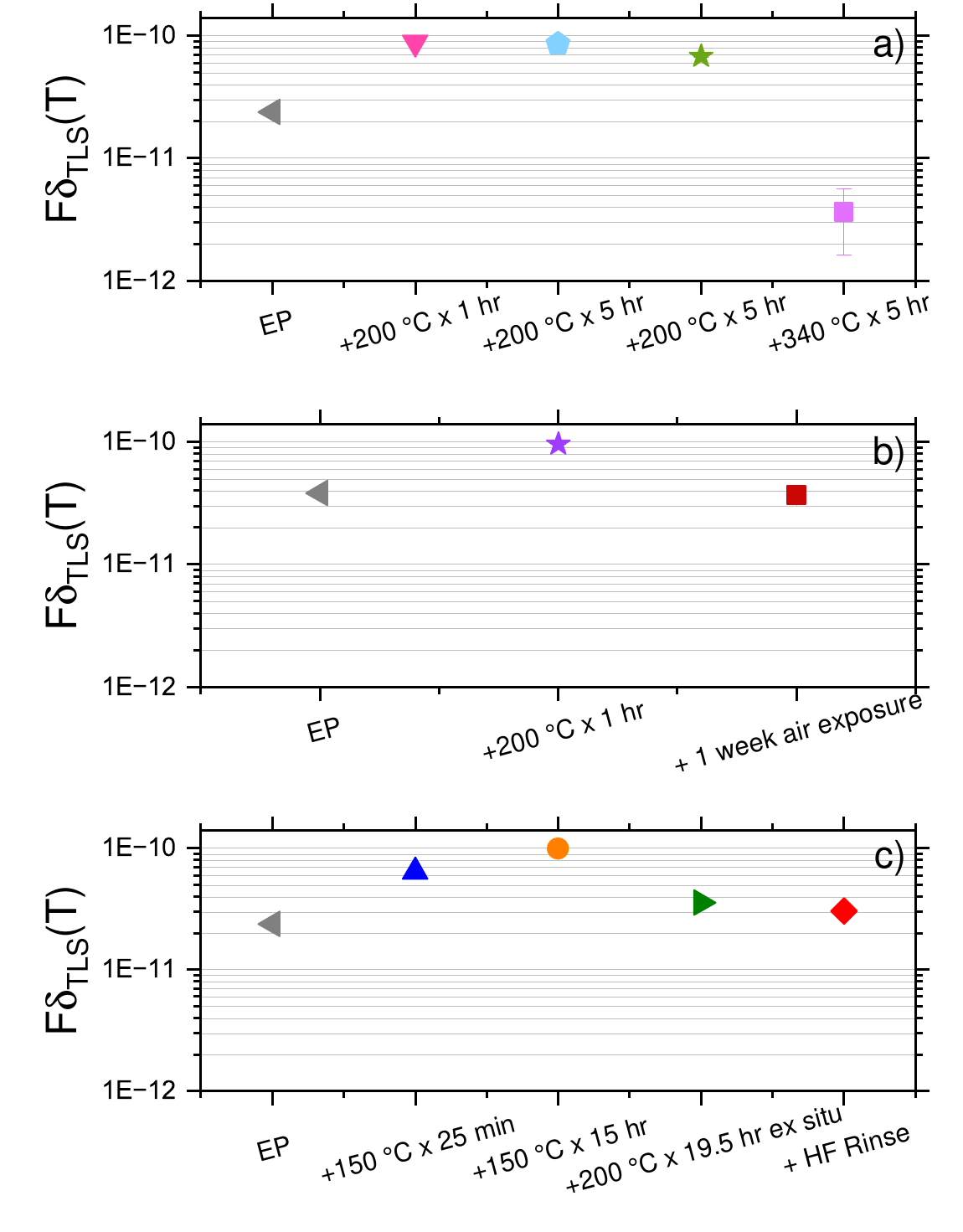}
\caption{Fitted $F \delta_{TLS}(T)$ values and fit error of a) TE1AES019, b) TE1RI003, and c) TE1AES021 using the data presented in Fig.~\ref{fig:QvsE} for 1.4~K~$<T<$~1.6~K. The 90$^{\circ}$C~$\times$~384~hours \textit{in situ} vacuum bake that precedes the sequential \textit{in situ} vacuum baking treatments for resonator TE1AES019 has been omitted from the horizontal axis for brevity. Most error bars are smaller than the symbol size.}
    \label{fig:LossTangent}
\end{figure}

We fit the data in Fig.~\ref{fig:QvsE} with the standard TLS model with a quasi-particle term built in. The electric field dependence of a niobium SRF resonator at low fields may be modelled as a single interface system (metal/air) containing TLS according to

\begin{equation}
    \frac{1}{Q_0} = \frac{F \delta _{TLS}(T)}{\big(1+\big(\frac{E_{acc}}{E_c}\big)^{2}\big)^{\beta}} + \frac{1}{Q_{qp}}
    \label{eq:PowDepTLS}
\end{equation}

\noindent where $\beta$ is a fitting parameter, $E_{c}$ is the characteristic saturation field, $\delta_{TLS}$(T) is the temperature dependent TLS dielectric loss tangent, and $Q_{qp}$ is the quasi-particle contribution to RF loss \cite{Martinis_PRL_2005, Rom17}. Fig.~\ref{fig:LossTangent} plots the product of the filling factor and dielectric loss tangent for each curve presented in Fig.~\ref{fig:QvsE}. We choose to present this convolved product as it is agnostic to the precise values of the filling factor and loss tangent and gives a qualitative measure of the TLS contribution to RF loss. Resonator TE1AES019 shown in the upper panel of Fig.~\ref{fig:LossTangent} displays a non-monotonic dependence of the TLS-driven losses with \textit{in situ} vacuum baking, achieving as low as 3.63E-12 after the final round of treatment, consistent with previous results \cite{Rom20}. TE1RI003 presented in the middle panel of Fig.~\ref{fig:LossTangent} further suggests that \textit{in situ} vacuum baking increases TLS-driven RF losses while reoxidation in dry clean room air for one week returns them to baseline EP values. The results of TE1AES021 in the lower panel further support this observation and suggest that a fully oxidized surface, whether dry or wet grown, contains less two-level system loss than a partially dissolved oxide.

\begin{figure}
    \centering
    
    \includegraphics[width=8.6cm]{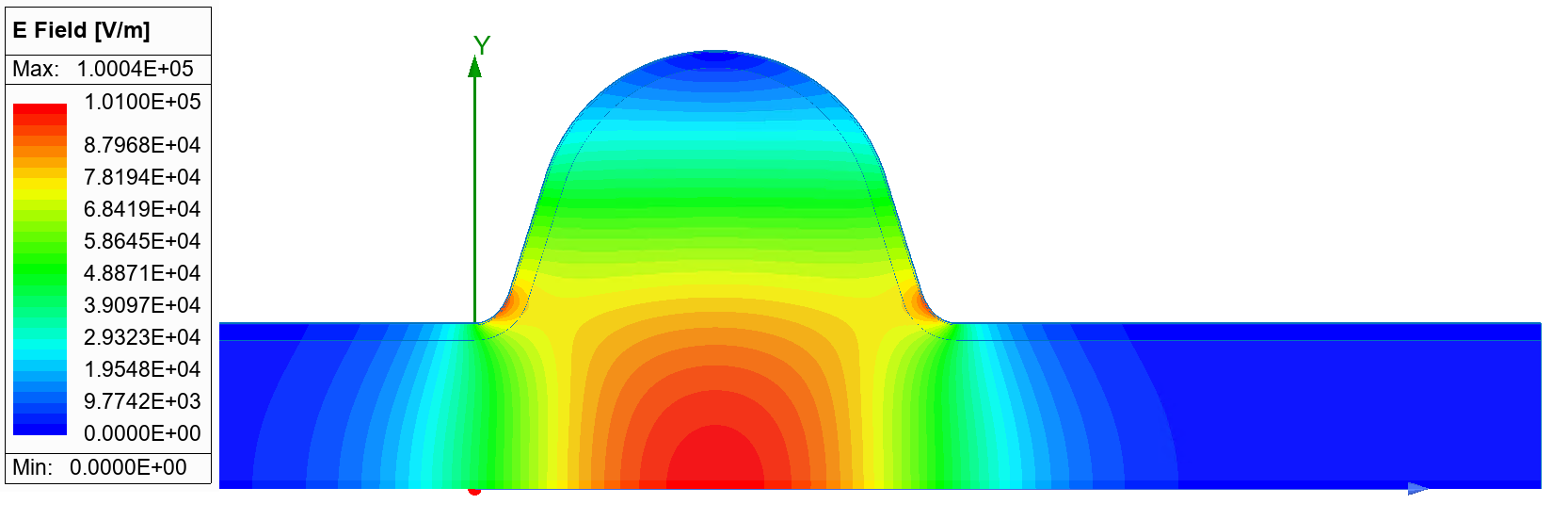}
    \caption{Electric field distribution with $E_{acc}$ $\approx$ 0.1 MV/m.}
    \label{fig:fieldDistribution}
\end{figure}

We mention that the characteristic saturation field extracted from Fig.~\ref{fig:QvsE} remains roughly constant, as evidenced by the increase in $Q_0$ at 0.1-0.2~MV/m for all investigated cavities and temperatures. This supports the theory that a single loss mechanism remains dominant throughout these studies: two-level systems located within the oxide. However, such a critical field is considerably higher than the values of $\sim$1~V/m observed at mK in 2-D devices \cite{Sarabi_thesis,Sarabi_PRL_2016}. We believe that this is due to two factors. First, the large loss tangent of the pentoxide \cite{Rom20} may require a large electric field (i.e. more photons) to saturate the TLS. Moreover, due to the inhomogeneous electric field distribution within the cavities, when the reported on-axis field is $E_{acc} \approx$~0.1~MV/m, surface TLS interact with electric fields which vary from $\sim$0 - 8e5 V/m, as depicted in Fig.~\ref{fig:fieldDistribution}. Second, the data reported in Fig.~\ref{fig:QvsE} was obtained at comparatively higher temperatures of 1.4~K-1.6~K; it has been theoretically and experimentally shown that $E_c$ increases with increasing temperature \cite{Gao_thesis}. As a result, the values of $E_c$ are in line with what might be expected from the conventional theory of TLS.

\subsection{ToF-SIMS studies}

\begin{figure}[h!]
    \centering
    \includegraphics[width=8.6cm]{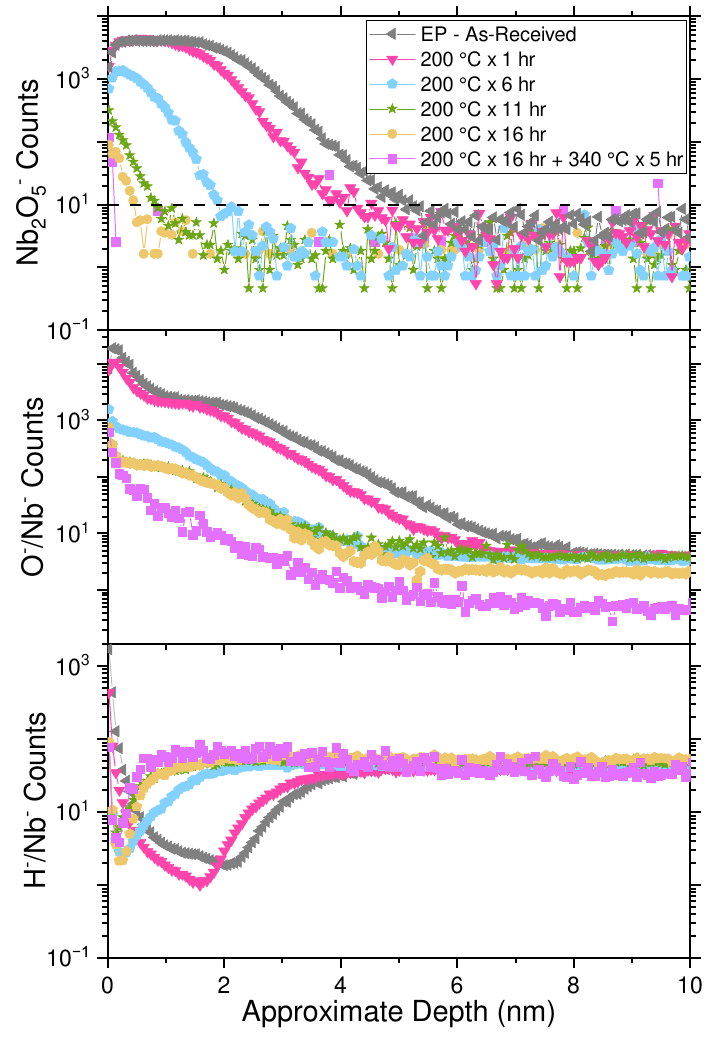}
    \caption{Averaged ToF-SIMS depth profiles of the Nb\textsubscript{2}O\textsubscript{5}\textsuperscript{-}, O\textsuperscript{-}/Nb\textsuperscript{-}, and H\textsuperscript{-}/Nb\textsuperscript{-} signals acquired on an as-received electropolished resonator cutout subjected to sequential rounds of \textit{in situ} vacuum baking. The Nb\textsubscript{2}O\textsubscript{5}\textsuperscript{-} signals are scaled relative to one another such that the values at 10~nm are identical. Nb\textsubscript{2}O\textsubscript{5} thickness $t$ is defined as the depth at which counts drop below 10, as denoted by the dashed line. Durations presented in the legend represent integrated bake times.}
    \label{fig:sims}
\end{figure}

As the amorphous niobium oxide is known to host TLS, it is likely that its dynamics dictate the performance evolution observed in Figs.~\ref{fig:QvsE} and \ref{fig:LossTangent}. To investigate this possibility, we recreated the \textit{in situ} baking treatments presented in Fig.~\ref{fig:QvsE}a) on an electropolished niobium SRF resonator cutout in the preparation chamber associated with a ToF-SIMS setup. This allowed us to perform treatments and measure changes in the surface chemistry without re-exposing to ambient conditions. We used a liquid bismuth ion Bi\textsuperscript{+} beam to perform secondary ion mass measurements while sputtering with a 500~eV cesium ion Cs\textsuperscript{+} gun to obtain depth profiles of various species. The sample was sequentially vacuum baked in steps of 200$^{\circ}$C followed by cool down to and measurement at room temperature \textit{in situ}. The final bake was performed at 340$^{\circ}$C for 5 hours. Three depth profiles were obtained at different spots on the sample after each step of treatment to ensure sample uniformity. All measurements were performed in a vacuum better than 1E-10~Torr. Fig.~\ref{fig:sims} presents the resulting averaged Nb\textsubscript{2}O\textsubscript{5}\textsuperscript{-}, O\textsuperscript{-}/Nb\textsuperscript{-}, and H\textsuperscript{-}/Nb\textsuperscript{-} signals as a function of depth into the samples.

The as-received electropolished resonator cutout displays a fully intact 5~nm thick native pentoxide. After the first round of \textit{in situ} baking at 200$^{\circ}$C for 1~hour, we find that this Nb\textsubscript{2}O\textsubscript{5} layer is reduced slightly, extending to a thickness of 4~nm. Continued baking shows further reduction in pentoxide thickness and is eliminated by the 340$^{\circ}$C~$\times$~5~hour bake. These results confirm the gradual dissolution of the native Nb\textsubscript{2}O\textsubscript{5} layer with sequential \textit{in situ} vacuum baking at 200$^{\circ}$C and the subsequent inward diffusion of oxygen over the oxide/metal interface  \cite{Ma_JAP_2004, Schulze_1977, Fromm_1977, Del_APL_2008, Lechner_APL_2021, Bafia_SRF_2021}. Coincident with the dissolution of the oxide is the decrease of the H\textsuperscript{-} signal within it. We note, however, that the H\textsuperscript{-} signal in the niobium metal below the oxide remains unchanged, suggesting that the concentration of hydrogen in the niobium metal remains constant throughout this study.

\begin{figure}
    \centering
    \includegraphics[width=8.6cm]{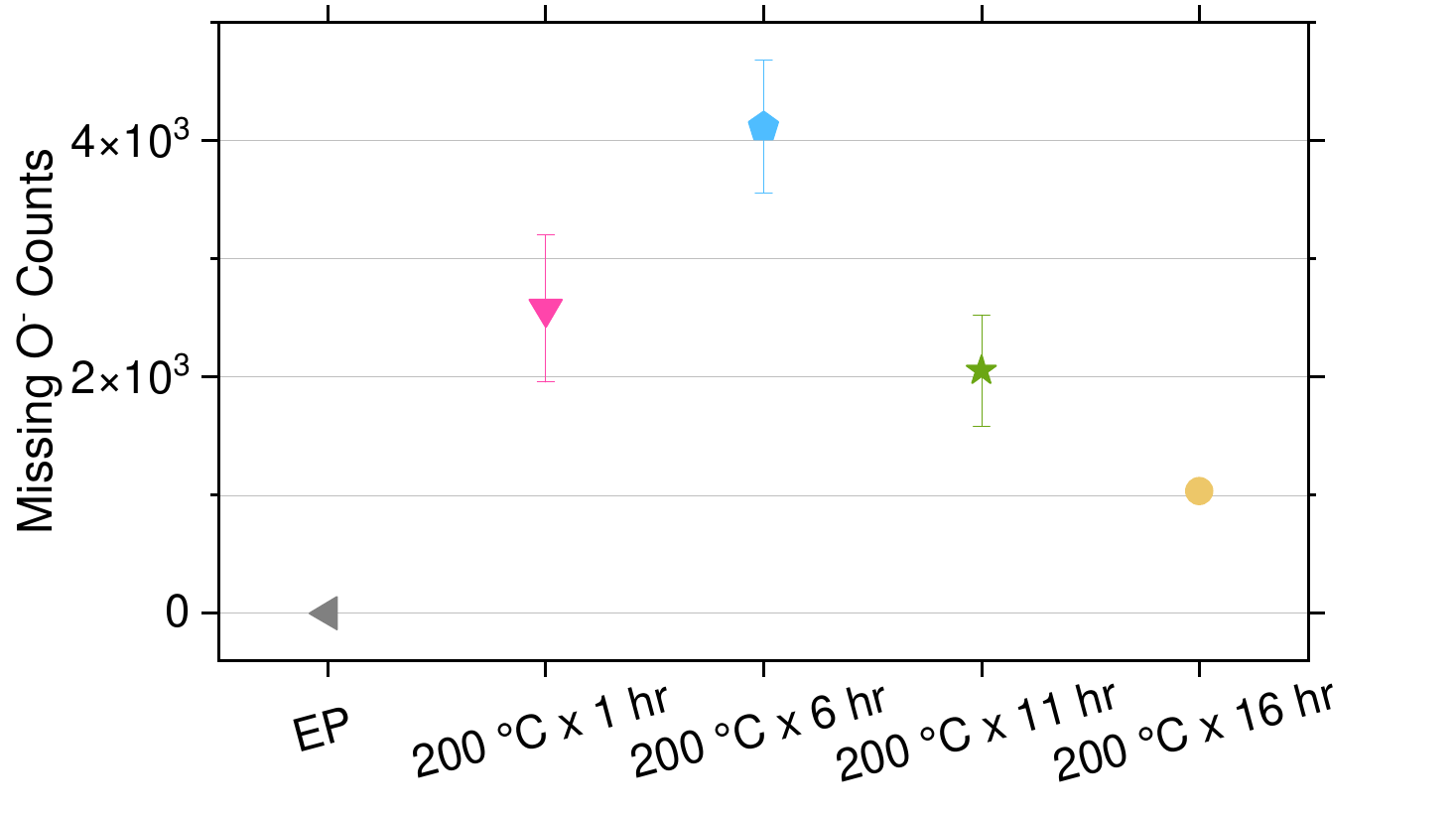}
    \caption{Averaged missing oxygen counts in the niobium oxide obtained using data presented in Fig.~\ref{fig:sims}. These values are proportional to the number of oxygen vacancies in the oxide; see text for details. Error bars depict standard deviation.}
    \label{fig:MissingOCount}
\end{figure}

As discussed above, oxygen vacancies in the niobium pentoxide may serve as a potential source of TLS-driven losses in niobium-based resonators; to investigate if such vacancies correlate with the evolution of our cavity performance, we extract out the number of missing oxygen atoms within the pentoxide layer for each treatment presented in Fig.~\ref{fig:sims}. This is done by first scaling the total oxygen counts by the pentoxide thickness to obtain the expected oxygen counts per unit thickness $\sigma$ for the as-received, EP baseline sample. Second, for each annealing step, we calculate the \textit{expected} oxygen counts by scaling $\sigma$ with the measured pentoxide thickness $t$. We then calculate the measured oxygen counts $o$ from the O\textsuperscript{-}/Nb\textsuperscript{-} signal in Fig.~\ref{fig:sims} by integrating from the surface to depth $t$. Lastly, we take the difference between the expected and actual measured oxygen counts such that we have $\sigma t - o$. The result yields the missing oxygen counts in the pentoxide which is proportional to the total oxygen vacancies in the sample. We note that this analysis is not applicable to the results post 340$^{\circ}$C~$\times$~5~hour \textit{in situ} baking as the pentoxide is eliminated, making it difficult to quantify the number of missing counts in the absent layer. The results presented in Fig.~\ref{fig:MissingOCount} show a non-monotonic evolution with subsequent baking treatments. Such behavior could be explained by considering an interplay of oxygen vacancy generation and overall oxide thickness reduction. At first, baking produces many vacancies while the overall thickness decreases only slightly. However, as baking continues, the oxide thickness decreases dramatically, and the total number of vacancies within the oxide begins to decrease.

\section{Discussion\label{sec:Discussion}}

The non-monotonic increase and subsequent decrease in TLS-driven RF losses combined with the non-monotonic evolution of the number of missing oxygen counts suggests the role of oxygen vacancies in the pentoxide as a source of TLS. Indeed, previous tunneling and XPS measurements on cavity-grade niobium baked at various temperatures and durations show that, when compared to an unbaked niobium sample, the 250$^{\circ}$C~$\times$~2~hr bake not only causes Nb\textsubscript{2}O\textsubscript{5} to reduce but also yields an increased level of subgap states which is likely driven by magnetic moments formed due to the presence of oxygen vacancies within the oxide \cite{Proslier_2008}. Moreover, other work has shown that these magnetic moments increase with oxygen vacancy concentration \cite{Cava_PRB_1991}. As such, \textit{in situ} vacuum baking may promote the proliferation of oxygen vacancies that drive TLS-like losses in 3-D niobium resonators. The aggravation and subsequent mitigation of TLS-driven losses post \textit{in situ} vacuum baking is likely due to the interplay of oxygen vacancy generation in the Nb\textsubscript{2}O\textsubscript{5} layer and overall pentoxide dissolution.

This hypothesis is further corroborated by the fact that treatments which re-expose the resonator to air and reform the native oxide show nearly identical TLS-driven losses when compared to the baseline EP test. Moreover, despite the fact that TE1RI003 post re-oxidation and TE1AES021 post the additional HF acid rinse contained a larger concentration of interstitial oxygen in the underlying niobium material, both gave nearly identical performance to that of the baseline EP values in Figs.~\ref{fig:QvsE} and \ref{fig:LossTangent}. This localizes the TLS losses to the oxide and asserts the non-role of interstitial oxygen in niobium at these fields and temperatures. This further supports the conclusions made in \cite{Rom17}. 

We find that hydrogen is likely not responsible for the variations in performance that we observe here. From Fig.~\ref{fig:sims}, we find that the hydrogen concentration at the surface decreases monotonically. Moreover, the hydrogen in the niobium metal does not change appreciably with the baking. Neither of these experimental observations can be used to explain the non-monotonic evolution of the RF performance.

\begin{figure}
    \centering
    \includegraphics[width=7.6cm]{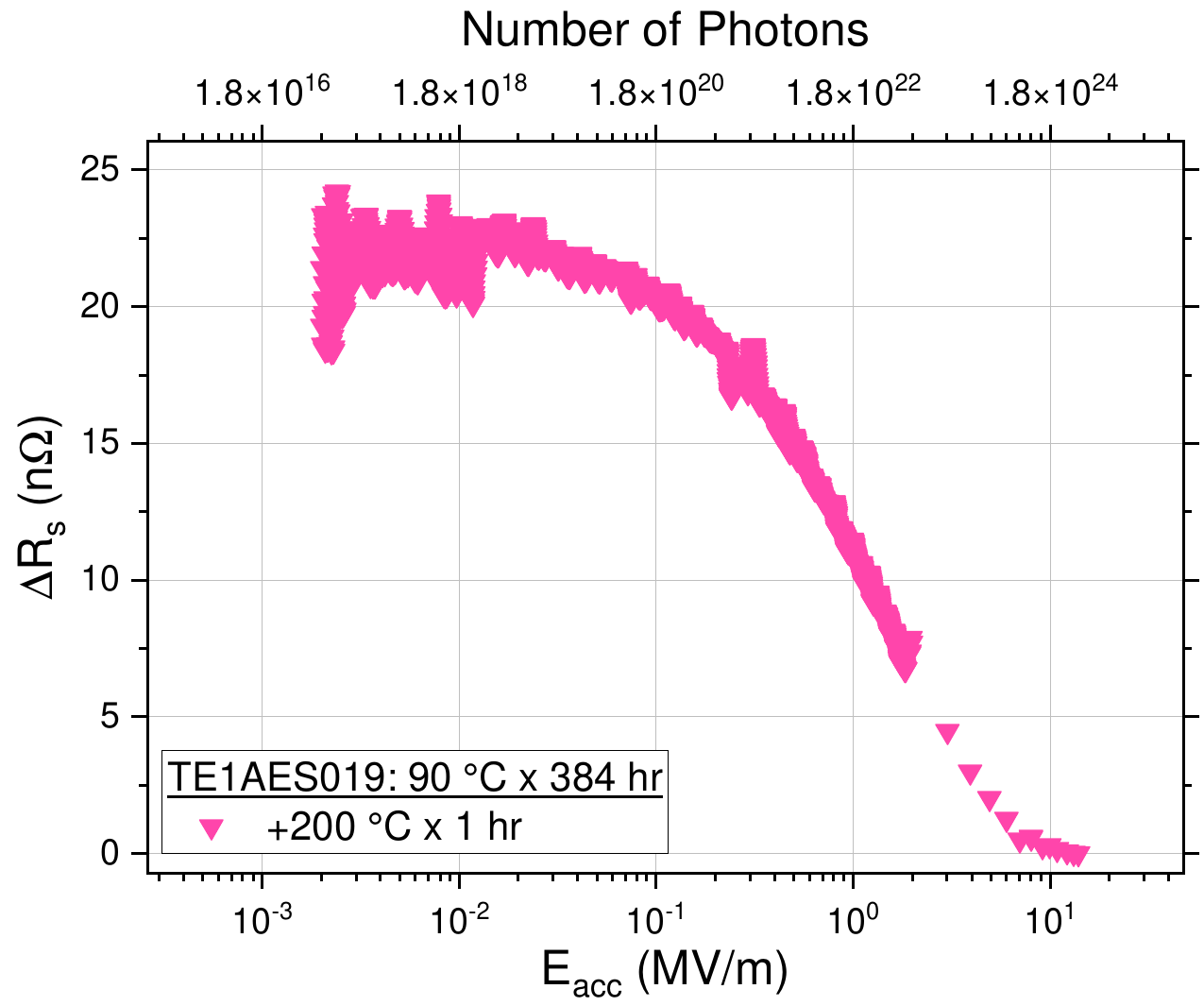}
    \caption{Surface resistance introduced by TLS for resonator TE1AES019 post 90$^{\circ}$C~$\times$~384~hour + 200$^{\circ}$C~$\times$~1~hour shown in the upper panel in Fig.~\ref{fig:QvsE}.}
    \label{fig:AES19_Rs}
\end{figure}

Fig.~\ref{fig:AES19_Rs} presents the additional surface resistance due to TLS-induced dissipation for resonator TE1AES019 post the 90$^{\circ}$C~$\times$~384~hours~$+$~200$^{\circ}$C~$\times$~1~hour vacuum baking treatment. This was obtained using the relation $R_s = G/Q_0$ ($G$ is a known geometry factor) and by subtracting off the contribution from non-TLS losses. Surprisingly, the surface resistance at low fields levels off at 22.5~n$\Omega$, roughly two times larger than what was reported for a 100~nm thick niobium oxide grown \textit{via} anodization \cite{Rom17}. This is also well explained by the Nb\textsubscript{2}O\textsubscript{5} vacancy hypothesis as oxides grown through anodization are known to contain fewer oxygen vacancies than wet grown oxides \cite{Grund1980}. As a result, while the partially depleted oxide post vacuum baking is significantly thinner than 100~nm, it may host a larger concentration of magnetic moments within the pentoxide and thus introduce a greater level of TLS losses. 

While the present work focuses on the effect of vacuum baking on 3-D niobium SRF resonator performance in the quantum regime, the observations made here are applicable to transmon qubits where niobium oxides are present. As many qubit fabrication processes perform vacuum baking treatments after the deposition and oxidation of niobium thin films, there exists the potential of generating a magnetic pentoxide which may further limit transmon qubit quantum coherence times. 

\section{Conclusion\label{sec:Conclusion}}

In summary, we studied the effect of low temperature vacuum baking on the quality factor of three 1.3~GHz 3-D niobium SRF resonators and found evidence which suggests that oxygen vacancies in the niobium pentoxide are a potential source of TLS losses that limit performance in both Nb resonators and transmon qubits which utilize oxidized Nb. \textit{In situ} vacuum baking produces an aggravation and subsequent mitigation of these losses due to the interplay of oxygen vacancy generation and oxide dissolution. 

\section{Acknowledgments}

This material is based upon work supported by the U.S. Department of Energy, Office of Science, National Quantum Information Science Research Centers, Superconducting Quantum Materials and Systems Center (SQMS) under contract number DE-AC02-07CH11359.

\bibliography{Main}
\end{document}